\documentstyle[aps,prl]{revtex}

\begin{document}
\draft
\title{Resonance phenomena of a soliton-like extended object in a bistable potential}
\author{J. A. Gonz\'alez, B. A. Mello, L. I. Reyes and L. E. Guerrero}
\address{Centro de F\'{\i}sica, Instituto Venezolano de Investigaciones\\
Cient\'{\i}ficas, Apartado Postal 21827, Caracas 1020-A, Venezuela}
\date{\today}
\maketitle

\begin{abstract}
We investigate the dynamics of a soliton that behaves as an extended
particle. The soliton motion in an effective bistable potential can be
chaotic in a similar way as the Duffing oscillator. We generalize the
concept of geometrical resonance to spatiotemporal systems and apply it to
design a nonfeedback mechanism of chaos control using localized
perturbations. We show the existence of {\it solitonic stochastic resonance}.
\end{abstract}


\preprint{HEP/123-qed}

\ 


The soliton dynamics in inhomogeneous media \cite{Kivshar91}-\cite{Mello},
the spatiotemporal chaos \cite{Gonzalez92}, \cite{Willeboordse}-\cite
{Guerrero89}, the chaos control \cite{Ott}-\cite{Anerbach} and diverse types
of resonance phenomena \cite{Benzi}-\cite{Chacon96l} have become the object
of intensive study in recent years (although rarely these phenomena have
been studied simultaneously).

In the present letter we investigate a solitonic model perturbed by
inhomogeneous forces \cite{Gonzalez92}-\cite{Mello}, for which the exact
solution describing the stationary equilibrium solitonic state can be
obtained. Furthermore, the stability problem for this solution can be solved
exactly. The effective potential for the soliton motion can be bistable
depending on the system parameters. Using an additional external periodic
force we can have the soliton jumping chaotically between the two potential
wells (like in the Duffing oscillator) \cite{Guckenheimer}. We prove that
changes of the soliton shape and the wave-form of the external perturbations
are very relevant to the soliton's dynamics as a whole. We generalize the
concept of geometrical resonance \cite{Chacon96l} to spatiotemporal systems.
This concept can be used to design a nonfeedback mechanism of chaos control 
\cite{Braiman}-\cite{Qu}. In our case, this mechanism can be applied in a
localized way in space. Finally, we show the existence of solitonic
stochastic resonance (SSR).

We are interested in equations of the type: 
\begin{eqnarray}
\phi _{tt} &-&\phi _{xx}+R(\phi ,\phi _t,x)-G(\phi )  \nonumber \\
\ &=&F(x)+P(\phi ,\phi _t)q(x,t),  \label{EqMov}
\end{eqnarray}
where $G(\phi )=-\frac{\partial U(\phi )}{\partial \phi }$; $U(\phi )$ is a
potential that possesses at least two minima \cite{Gonzalez87}, \cite
{Gonzalez89}; $R(\phi ,\phi _t,x)$ is a dissipative term, which in general
can be nonlinear \cite{Guerrero96} and the damping coefficient can depend on 
$x$; $F(x)$ is an inhomogeneous force, and $P(\phi ,\phi _t)q(x,t)$ is a
general temporal perturbation. Many systems are described by this type of
equations, including charge density waves, Josephson junctions, and
structural phase transitions \cite{Bishop}.

Problems related to Eq. (\ref{EqMov}) present extraordinary mathematical
difficulties. Even if we are to solve (for case $P(\phi ,\phi
_t)q(x,t)\equiv 0$) the stability problem of a stationary soliton placed on
an equilibrium position created by the inhomogeneities \cite{Gonzalez92}-%
\cite{Mello}, this task requires the making of big approximations \cite
{Kivshar89}. The most common approximation is to consider the soliton as
structureless point-like particle \cite{Kivshar89} (in this case the
coordinate of the soliton center-of-mass and its velocity are the only
dynamical variables). Although this is a valid approximation in many cases,
our work \cite{Gonzalez92}-\cite{Mello} has shown that, in general, the
internal dynamics plays a fundamental role.

Besides the applications of the $\phi ^4$-equation in phase transition theory%
\cite{Bishop}, it is a {\it model system} for topological solitons in
general. Let us consider as an example the perturbed $\phi ^4$-model: 
\begin{equation}
\phi _{tt}-\phi _{xx}+\gamma \phi _t-\frac 12(\phi -\phi ^3)=F(x),
\label{PerPhi4}
\end{equation}
where 
\begin{eqnarray}
F(x)=F_1(x) &\equiv &\frac 12A(A^2-1)\tanh (Bx)+  \nonumber \\
&&\ \frac 12A(4B^2-A^2)\frac{\sinh (Bx)}{\cosh ^3(Bx)}.  \nonumber
\end{eqnarray}

The inhomogeneities are chosen in order to have some important properties.
The exact stationary solution for the soliton equilibrated by the
inhomogeneities in the point $x=0$ can be obtained $\phi _k=A\tanh (Bx).$
The stability problem of this solution can be solved exactly. The force $%
F_1(x)$ possesses the interesting property to be topologically equivalent
(in the sense of Catastrophe Theory \cite{Zeeman}) to a pitchfork
bifurcation, allowing us to have an effective potential (for the soliton)
with one or two wells, depending on the system parameters. This system is
generic and structurally stable. Therefore the results obtained in this
letter can be generalized qualitatively to other systems topologically
equivalent to that described by Eq.\ref{PerPhi4}. The force $F(x)$ permits
to observe important phenomena which occur with the participation of the
internal soliton dynamics including the appearance of a great number of
internal modes and the soliton destruction due to the instability of the
shape mode in some special situations \cite{Gonzalez92}-\cite{Mello}.

The stability analysis \cite{Gonzalez92}-\cite{Mello} which considers small
perturbations around the soliton ($\phi (x,t)=\phi _k(x)+f(x)e^{\lambda t}$)
leads to the eigenvalue problem $\hat Lf=\Gamma f,$ where $\hat L=-\partial
_x^2+\left( \frac 32A^2-\frac 12-\frac{3A^2}{2\cosh ^2(Bx)}\right) $, $%
\Gamma =-(\lambda ^2+\gamma \lambda )$.

The eigenvalues of the discrete spectrum are given by the expression $\Gamma
_n=-\frac 12+B^2(\Lambda +2\Lambda n-n^2),$ where $\Lambda (\Lambda +1)=%
\frac{3A^2}{2B^2}$. The stability condition of the equilibrium point $x=0$
(for the soliton) is defined by the eigenvalue of the translational mode ($%
f_0(x)=\cosh ^{-\Lambda }(Bx)$): $B^2\Lambda -\frac 12>0.$ A global
topological analysis and the investigation of the infinity \cite{Mello}
allow us to have a complete qualitative information of the soliton dynamics.
When the equilibrium position for the soliton is unstable and the absolute
value of the eigenvalue corresponding to the translational mode is
sufficiently high, the first shape mode ($f_1(x)=\sinh (Bx)\cosh ^{-\Lambda
}(Bx)$), and even other internal modes, can be unstable too, producing the
soliton destruction \cite{Mello}.
We have at least two interesting situations: $\mbox{If }A^2>1\mbox{ and }%
2\Lambda B^2>1,$ then, there exists only one stable equilibrium point for
the soliton ($x_{00}=0$). $\mbox{If }A^2>1\mbox{ and }2\Lambda B^2<1,$ a
pitchfork-like bifurcation occurs and we have two stable equilibrium points (%
$x_{01}<0$, $x_{02}>0$) (the point $x_{00}=0$ becomes unstable). It is
important to stress that the bifurcation does not occur at the point we
would expect when considering the soliton as a structureless point particle
(in this case the number of equilibrium positions depend on the number of
zeroes of force $F(x)$).

If the system, in the bistable case, is perturbed by an additional periodic
force fitted to the shape of the translational mode\cite{Gonzalez92},\cite
{Guerrero96} (in (Eq.\ref{EqMov}) we put $P(\phi ,\phi _t)\equiv 1$), then
we can have similar situations to those of the Duffing oscillator \cite
{Guckenheimer}. But remember that in this case what is jumping between the
potential wells is an extended object with a very complicated internal
dynamics. On the other hand, for sufficiently high values of $|\Gamma _0|$
for the unstable position, the soliton can bifurcate \cite{Mello} in an
antisoliton (which would feel the middle position as a stable one), and two
solitons, each of which would move towards one of the wells.

The recently formulated concept of geometrical resonance \cite{Chacon96l}
(GR) can be very useful in this context. Generalizing this concept for
spatiotemporal systems we define $\phi _{GR}(x,t)$ as a GR solution of
equation (\ref{EqMov}) if it satisfies the condition 
\begin{equation}
R(\phi _{GR},\partial _t\phi _{GR},x)=P(\phi _{GR},\partial _t\phi
_{GR})q_{GR}(x,t).  \label{GRCon}
\end{equation}
In this case there exists (local) conservation of energy: 
\begin{eqnarray}
H\equiv \int_{-\infty }^\infty &&\left[ \frac 12\left( \frac{\partial \phi
_{GR}}{\partial t}\right) ^2+\frac 12\left( \frac{\partial \phi _{GR}}{%
\partial x}\right) ^2+\right.  \nonumber \\
&&\Biggl. U(\phi _{GR})-F(x)\phi _{GR}+C\ \Biggr] \mbox{dx}
\end{eqnarray}
If we want to observe a GR situation, we should use a $q_{GR}(x,t)$ such
that it satisfies the GR condition (\ref{GRCon}).

Consider the Eq. (\ref{EqMov}) with $R=\gamma \phi _t$; $P=1$; $G(\phi )=%
\frac 12(\phi -\phi ^3)$ and $F(x)$ as defined in (\ref{PerPhi4}).

We assume that the stability condition $B^2\Lambda -\frac 12>0$ for the
equilibrium position $x=0$, holds. In this case we can write an approximate
solution for $\phi (x,t)$ ($q\equiv 0$): 
\begin{equation}
\phi (x,t)=A\tanh (Bx)+\frac{h_{00}\cos (\omega _0t+\theta _0)}{\cosh
^\Lambda (Bx)},  \label{AprSol}
\end{equation}
where $\omega _0^2=\Gamma _0$.

Thus, for small values of $h_{00}$ the perturbation 
\begin{equation}
q_{GR}(x,t)=-\frac{\omega _0h_{00}\gamma \sin (\omega _0t+\theta _0)}{%
cosh^\Lambda (Bx)}  \label{Per}
\end{equation}
satisfies the GR condition. This explains a whole series of experiments
preformed in \cite{Mello}, \cite{Guerrero96}. There the $\phi ^4$-kink
confined in a potential well created by the inhomogeneity $F(x)$ was
perturbed by the time-dependent force (\ref{Per}). The authors have got
resonances of the translational mode practically without deformation of the
kink profile. Chaotic behavior is obtained only for high values of the
perturbation amplitude, when (\ref{AprSol}) is not a GR solution anymore.
Meanwhile, if we use a different time-dependent force (e.g., $q(x,t)=\rho
_0\cos \omega t$) which does not satisfy the GR condition, we will obtain
irregular behavior in time and space with much smaller amplitudes.

The GR \cite{Chacon96l} provides a mechanism for nonfeedback control of
chaos.

Suppose we have the following equation 
\begin{eqnarray}
&&\phi _{tt}-\phi _{xx}+\gamma \phi _t-\frac 12(\phi -\phi ^3)  \nonumber \\
&=&F(x)-P_0\frac{\cos (\omega t)}{\cosh ^2(Bx)}+F_c(x,t).  \label{EqCon}
\end{eqnarray}
We assume that for certain initial conditions the Eq. (\ref{EqCon}) is in a
chaotic regime provided $F_c(x,t)\equiv 0$.

We are left with the problem of eliminating the chaotic motion using control 
$F_c(x,t)$. Additionally, it is expected that the control driving term is
small and localized in space. In order to do this we should select the
control term in such a way that the solution will be sufficiently close to a 
$T^{\prime}$ periodic GR solution.

We can choose the control function in the form 
\begin{equation}
F_c(x,t)=\frac{g_c\cos (\omega _ct+\theta _c)}{\cosh ^\Lambda (Bx)}.
\label{ConFun}
\end{equation}
Using the concept of ``almost adiabatic invariant'' \cite{Arnold},\cite
{Chacon96l}, we arrive at a condition for the control parameters 
\begin{eqnarray}
&&\left\langle \int_{-\infty }^\infty \right. -\gamma \left( \frac{\partial
\phi _{GR}}{\partial t}\right) ^2+P_0\frac{\partial \phi _{GR}}{\partial t}%
\cos (\omega t)+  \nonumber \\
&&\left. g_c\frac{\partial \phi _{GR}}{\partial t}\frac{\cos (\omega
_ct+\theta _c)}{\cosh ^2(Bx)}\ \mbox{d}x\right\rangle _{T^{\prime }}=0.
\end{eqnarray}

The fact that we have taken (\ref{ConFun}) associated with the translational
mode (which approximately satisfies the GR condition) always allows us to
find control parameters (with small $g_c$) in order to suppress chaos. An
Arnold-like tongue structure, similar to that observed in \cite{Chacon96l}
can be obtained here. In Fig. 1 we show the results of the application of
the localized nonfeedback control.

Note that the validity of the concept of GR as a mechanism for chaos control
is not limited to the case in which the unperturbed equation can be solved
exactly. Using some topological analysis it is possible to guess the
``shape'' of the GR solution and then, through the concept of ``almost
adiabatic invariant'' $\left( \left\langle \frac{dH\left[ \phi _{GR}\right] 
}{dt}\right\rangle _{T^{\prime }}\approx 0\right) $one can obtain the
conditions such that the solution will be close to a GR solution inside of a
``mode-locked'' tongue.

The system we have presented in this letter with a soliton-like extended
object moving in an effective bistable potential can be also useful for
studying other phenomena, e.g. the spatiotemporal stochastic resonance (SR)
for the motion of an extended state with complicated internal dynamics.

Consider the equation 
\begin{eqnarray}
\phi_{tt} - \phi_{xx} + \gamma\phi_t - \frac{1}{2}(\phi-\phi^3) = F(x) +
q(x,t),
\end{eqnarray}

where 
\begin{eqnarray}
F(x)=\cases{B(4B^2-1)\tanh[B(x+x_0)] &for $x<0$, \cr 0 &for $x=0$, \cr
B(4B^2-1) \tanh[B(x-x_0)] &for $x>0$; }  \label{fuerzax}
\end{eqnarray}
\begin{eqnarray}
q(x,t)=\cases{[P_0\cos(\omega t) + \eta(x,t) ] \cosh^{-2}[B(x+x_0)] \cr \ \
\ \mbox{ for } x<0, \cr [P_0\cos(\omega t) + \eta(x,t) ] \
\cosh^{-2}[B(x-x_0)] \cr \ \ \  \mbox{ for } x>0;\cr }  \label{fuerzaxt}
\end{eqnarray}
here $4B^2>1$. The forces $F(x)$ and $q(x,t)$ are defined by Eq. (\ref
{fuerzax},\ref{fuerzaxt}) in such a way that there are two equilibrium
points for the soliton and the motion in each well is very close to the GR
condition. In the absence of the white noise $\eta (x,t)$ ($<\eta (x,t)>=0$, 
$<\eta (x,t)\eta (x^{\prime },t^{\prime })>=2D\delta (x-x^{\prime })\delta
(t-t^{\prime })$), the periodic force above is unable to make the soliton to
jump between the wells.

When we switch on the noise, it is possible to observe a maximum in the
graph of the Signal-to-Noise-Ratio (SNR) versus $D$ (see Fig. 2). Here as
the ``signal'' we take the time series of the soliton center-of-mass
coordinate: $x_{c.m.}=\frac{\int_{-l/2}^{l/2}x\phi _x^2dx}{%
\int_{-l/2}^{l/2}\phi _x^2dx}$. The SNR is measured following the
traditional method \cite{PToday}.

The maximum synchronization is obtained with a signal associated with the
translational mode. These results allow to predict the optimum entertainment
of the noise by means of a periodic signal, not necessarily a simple
sinusoidal signal but a more complex spatiotemporal function. The stochastic
resonance will depend on the characteristic shape of the kink and the
wave-form of the perturbation (in time and space). A manifestation of this
fact is the following phenomenon: if instead of the translational mode in
perturbation (\ref{fuerzaxt}) we use the first shape mode, then we do not
observe SSR.

Even when $F(x)$ has three zeros, if we are in the presence of an extended
object, the bistability is not a sufficient condition for the stochastic
resonance. The extended object should ``feel'' the bistability and the
internal modes should be stable in the vecinity of all the equilibrium
points, including the central equilibrium point which is unstable for the
translational mode. On the other hand, if the eigenvalue $\left| \Gamma
_0\right| $ corresponding to the unstable equilibrium position in the
bistable potential is very high, then the first shape mode can be unstable
too and the soliton will bifurcate in two solitons and one antisoliton. This
is what occurs when the parameters in Eq. (\ref{fuerzax}) are such that $%
B^2(4B^2-1)\tanh (Bx_0)>\frac{23}{50}$. The center of mass of this
``three-particle structure'' is always oscillating around $x=0$ and
therefore, there is no SSR.

The spatiotemporal stochastic resonance has been considered in some recent
papers \cite{Marchesoni}, \cite{Vilar2}. In particular, Ref. \cite
{Marchesoni} deals with the $\phi ^4$-equation. However, the SSR presented
here is a completely new phenomenon: there the behavior of the field $\phi $
is taken for the signal and the important bistability is in the potential $%
U(\phi )$, whereas in our work the relevant signal is the time series for
the coordinate of the soliton's center of mass. In Ref. \cite{Marchesoni}
there is no inhomogeneous force $F(x)$. The spatiotemporal stochastic
resonance studied there does not depend on $F(x)$. In our case the potential 
$U(\phi )$ is important for the existence of the soliton solution\cite
{Gonzalez87}, but it is the bistability in $x$, created by $F(x)$ which is
the key for the SSR. And, as we have seen not for every $F(x)$ the SSR
exists.

In general, we prove that changes of the soliton shape are very relevant to
its dynamics as a whole. Additionally, the wave-form of the perturbation is
crucial for all the resonance phenomena including the nonfeedback mechanism
of the chaos control and the stochastic resonance. We believe that these
phenomena are relevant to other systems where there is an extended state
with a complicated internal dynamics moving in a nonlinear potential force.

\begin{figure}[tbp]
\caption{Phase portraits for the motion of the soliton center of mass in eq.
(\ref{EqCon}), $F(x)$ is defined as in eq. (\ref{PerPhi4}) ($B=1/2$). a)
Chaotic motion in the absence of control ($F_c(x,t)\equiv 0$), $\gamma=0.1$, 
$A=1.5$, $P_0=1$ and $\omega=0.65$. b) Periodic motion as a result of the
application of nonfeedback control, $\omega_c=0.065$, $g_c=0.35$ and $%
\theta_c=0$.}
\end{figure}

\begin{figure}[tbp]
\caption{Two aspects of the solitonic stochastic resonance: a) There is a
maximum of the SNR at a critical value of noise intensity. b)
Sinchronization of the stochastic jumps with the periodic perturbation
linked with translational mode of the soliton.($P_0=0.04$, $\omega=0.01$, $%
x_0=2.5$ and $B=0.7$).}
\end{figure}

\begin{figure}[tbp]
\caption{Spatiotemporal evolution of the kink-soliton at the stochastic
resonance. Note that there is a minimal deformation of the kink profile}
\end{figure}


\begin{references}
\bibitem{Kivshar91}  Y. S. Kivshar, Zhang Fei and L. V\'azquez, Phys. Rev.
Lett. {\bf 67}, 1177 (1991).

\bibitem{Kivshar89}  Y. S. Kivshar and B. A. Malomed, Rev. Mod. Phys. {\bf 61%
}, 763 (1989).

\bibitem{Sanchez}  A. S\'anchez and A. R. Bishop, Phys. Rev. E{\bf \ 49},
4603 (1994).

\bibitem{Gonzalez92}  J. A. Gonz\'alez and J. A. Ho\l yst, Phys. Rev. B{\bf %
\ 45}, 10338 (1992).

\bibitem{Mello}  J. A. Gonz\'alez and B. A. Mello, Phys. Scripta {\bf 54},
14 (1996); Phys. Lett. A{\bf \ 219}, 226 (1996).

\bibitem{Willeboordse}  F. H. Willeboordse and K. Kaneko, Physica D {\bf 86}%
, 428 (1995).

\bibitem{Petrov}  N. Petrov, S. Metens, P. Borckmans, G. Dewel, K.
Showaltev, Phys. Rev. Lett. {\bf 75}, 2895 (1995).

\bibitem{Chate}  H. Chat\'e, Physica D{\bf \ 86}, 238 (1995).

\bibitem{Guerrero96}  J. A. Gonz\'alez, L. E. Guerrero and A. Bellor\'\i n,
Phys. Rev. E{\bf \ 54}, 1265 (1996).

\bibitem{Guerrero89}  L. E. Guerrero, and M. Octavio, Phys. Rev. A{\bf \ 40}%
, 3371, (1989).

\bibitem{Ott}  E. Ott, C. Grebogi and J. A. Yorke, Phys. Rev. Lett. {\bf 64}%
, 1196 (1990).

\bibitem{Ditto}  W. L. Ditto, M. L. Spano and J. F. Lindner, Physica D{\bf \
86}, 198 (1995).

\bibitem{Braiman}  Y. Braiman and I. Goldhirsch, Phys. Rev. Lett. {\bf 66},
2545 (1991).

\bibitem{Qu}  Z. Qu, G. Hu, G. Yang and G. Qin, Phys. Rev. Lett. {\bf 74},
1736 (1995).

\bibitem{Anerbach}  D. Auerbach, C. Grebogi, E. Ott and J. Yorke, Phys. Rev.
Lett. {\bf 69}, 3479 (1992).

\bibitem{Benzi}  R. Benzi, A. Sutera and A. Vulpiani, J.Phys. A{\bf \ 14},
L453 (1981).

\bibitem{Lindner}  J. F. Lindner, B. K. Meadows, W. L. Ditto, M. E. Inchiosa
and A. R. Bulsara, Phys. Rev. Lett. {\bf 75}, 3 (1995).

\bibitem{PToday}  A. R. Bulsara and L. Gammaitoni, Physics Today {\bf 49},
39 (1996).

\bibitem{Marchesoni}  F. Marchesoni, L. Gammaitoni and A. R. Bulsara, Phys.
Rev. Lett. {\bf 76}, 2609 (1996).

\bibitem{Vilar1}  J. M. G. Vilar and J. M. Rub\'\i , Phys. Rev. Lett. {\bf 78%
}, 2882 (1997)

\bibitem{Vilar2}  J. M. G. Vilar and J. M. Rub\'\i , Phys. Rev. Lett. {\bf 78%
}, 2886 (1997).

\bibitem{Chacon96l}  R. Chac\'on, Phys. Rev. Lett. {\bf 77}, 482 (1996).

\bibitem{Guckenheimer}  J. Guckenheimer and P. J. Holmes, {\em Nonlinear
Oscillations, Dynamical Systems and Bifurcations of Vector Fields}
(Springer, Berlin, 1986).

\bibitem{Gonzalez87}  J. A. Gonz\'alez and J. A. Ho\l yst, Phys. Rev. B{\bf %
\ 35}, 3643 (1987).

\bibitem{Gonzalez89}  J. A. Gonz\'alez and J. Estrada-Sarlabous, Phys. Lett.
A {\bf 140}, 189 (1989).

\bibitem{Bishop}  A. R. Bishop, J. A. Krumhansl, S. E. Trullinger, Physica D 
{\bf 1}, 1 (1980), and references therein.

\bibitem{Zeeman}  E. C. Zeeman, {\it Catastrophe theory: Selected Papers}
(Addison-Wesley, Reading, Massachusetts, 1977).

\bibitem{Arnold}  V. I. Arnold, {\em Geometrical Methods in the Theory of
Ordinary Differential Equations} (Springer-Verlag, New York, 1989).
\end{references}
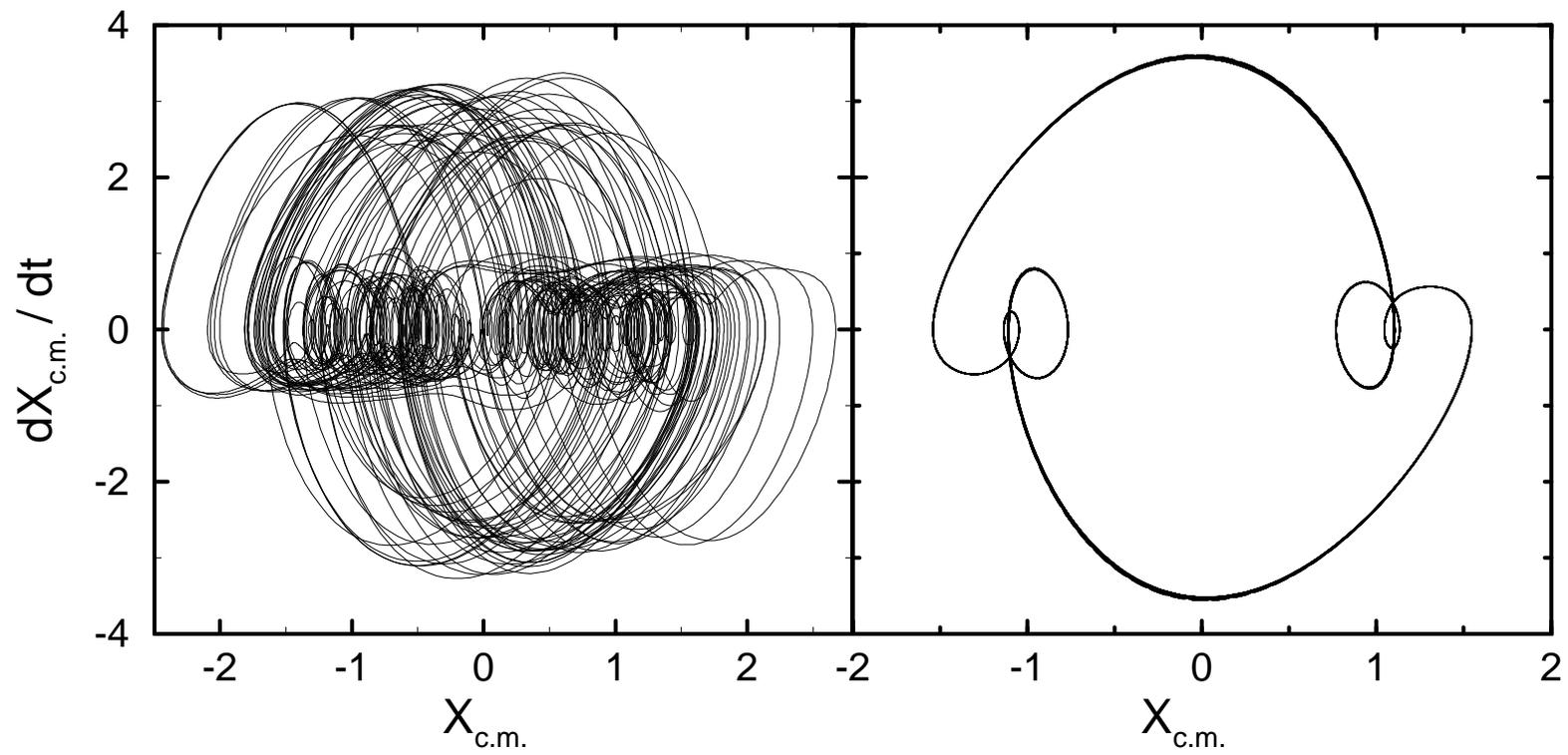
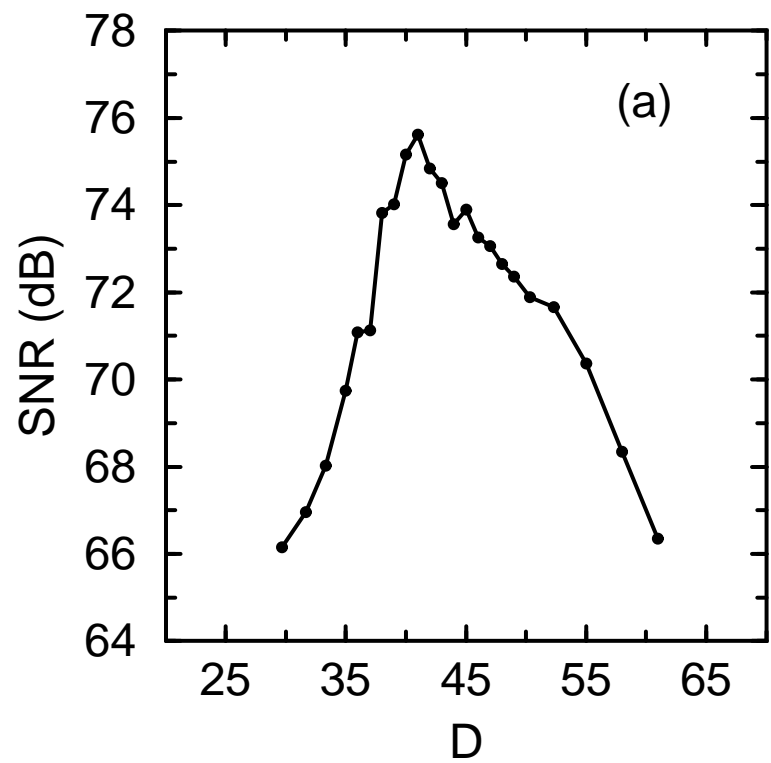
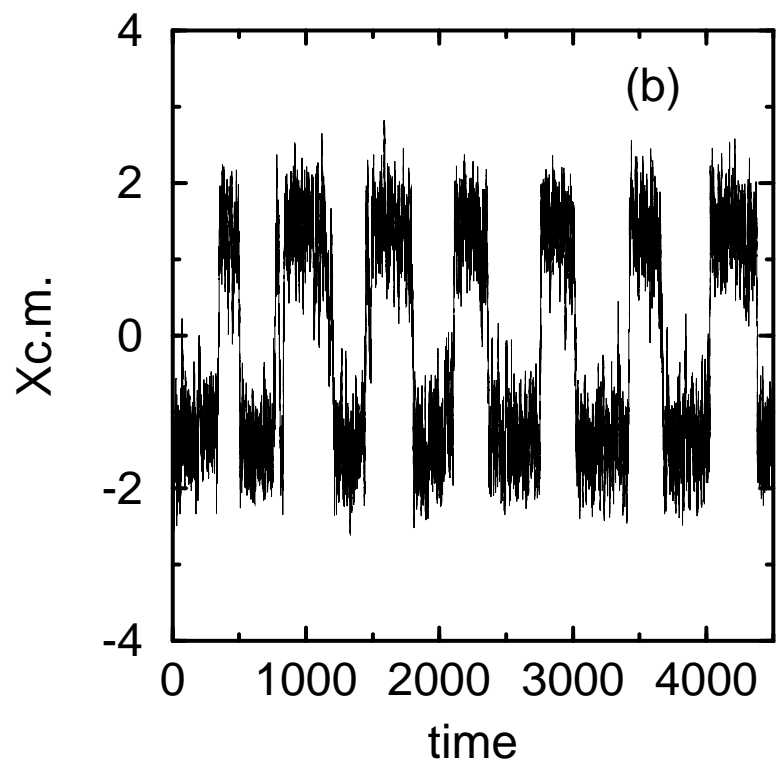
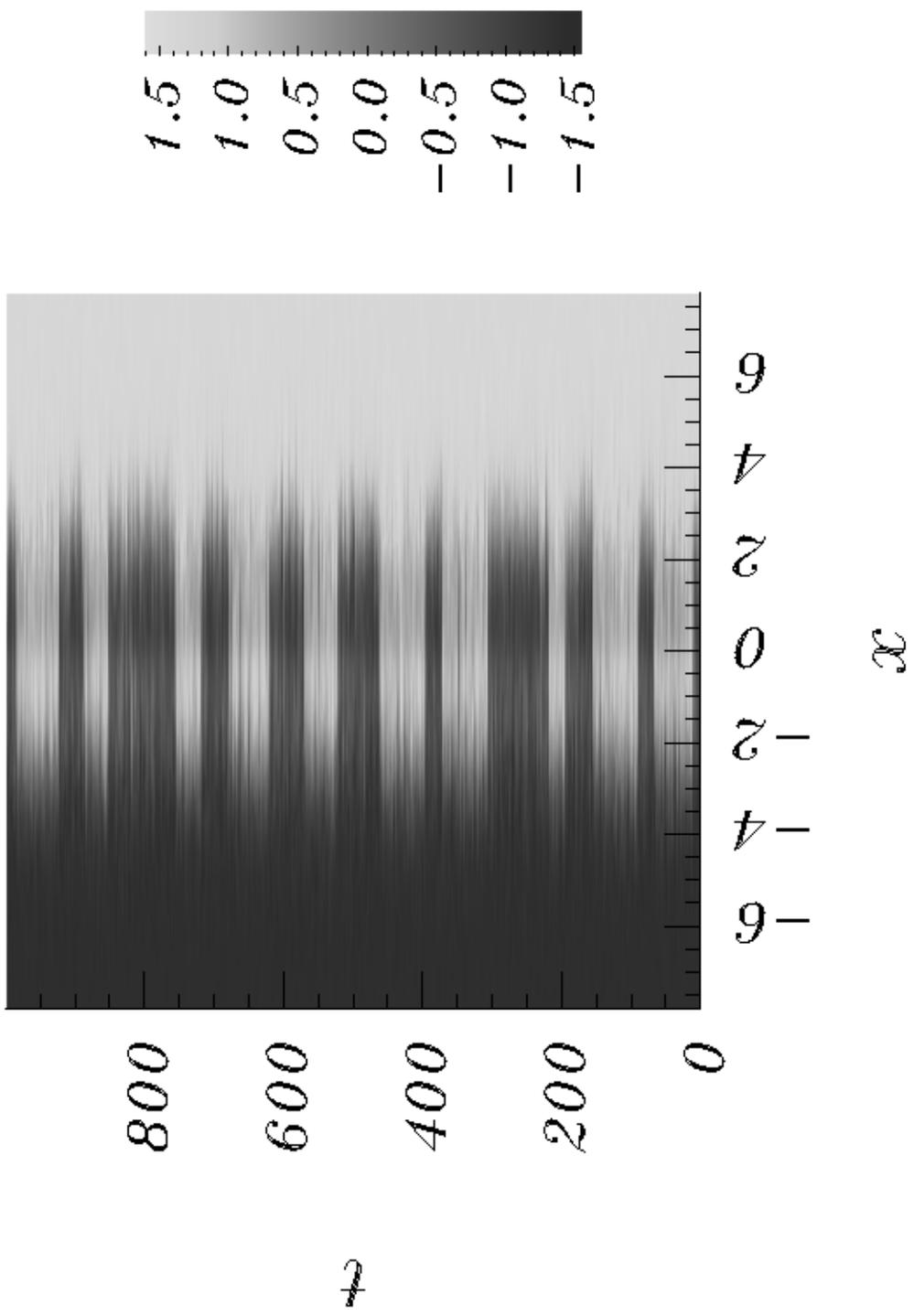
\end{document}